# Property Business Classification Model Based on Indonesia E-Commerce Data

Andry Alamsyah [1], Fariz Denada Sudrajat [2], Herry Irawan [3]

School of Economics and Business, Telkom University, Bandung, Indonesia

[1]andrya@telkomuniversity.ac.id, [2]farizdenada@gmail.com, [3]ir.herry@gmail.com

**Abstract**

*Online property business or known as e-commerce is currently experiencing an increase of home sales. Indonesia e-commerce property business has positive trending shown by the increasing sales for more than 500% from 2011 to 2015. A prediction of property price is important to help investor or public to have accurate information before buying property. One of methods for prediction is classification based on several distinctive of property industry attributes, such as building size, land size, number of rooms, and location.*

*Today, data is easily obtained, there are many open data from E-commerce sites. The E-commerce contains information about home and other property advertised to sell. People also regularly visit the site to find the right property or to sell the property using price information which collectively available as open data.*

*To predict the property sales, this research employed two different classification methods in Data Mining which are Decision Tree and k-NN classification. We compare which model classification is better to predict property price and their attributes. We use Indonesia biggest property based ecommerce sites Rumah123.com as our open data source, and choose location Bandung in our experiment. The accuracy result of decision tree is 75% and KNN is 71%, other than that k-NN can explore more data pattern than Decision Tree.*

## 1. Introductions

Property business currently is in positive trends in Indonesia, shown by the increasing number of transactions over the year. This opportunity force many property business company to develop their next year strategy. E-commerce is a way to generate new transactions by providing catalog for properties selling advertisement. The effective and efficient nature way doing property business transaction, makes e-commerce preferable for many Indonesian. The value of ecommerce in Indonesia is also predicted to become one of the best industry in the future [1].

Today, large collection of data can be easily obtained through internet. Those large-scale data contain many valuable information, if we can mine pattern or uncover hidden value using data analytics methodology. Data Analytics is the process of examining data sets to draw conclusions about the information. Data analytics techniques are widely used in commercial industries to enable organizations to make more-informed business decisions. One of data analytics methods is data mining. Some data have open nature which sometimes comes with no cost at all, for example data from online forum, social media, and e-commerce.

In this research, we explore open data from an e-commerce site to uncover hidden value of property business. Our result is a model to predict property price and their attributes. We collect data from e-commerce site *rumah123.com*, where it provides advertisement regarding property business. This site is the biggest e-commerce property business. We select one location which has growing property sales as case study, the location is *Bandung* City. This research describes patterns of the home sales market and potential place in Bandung by using *Decision Tree* and *k-Nearest Neighbor* classification model. The reason we use both model is because the suitability of ecommerce data with both model rather than other classification model.

## 2. Theoretical Background

Data mining is a process to discover insightful, interesting patterns, as well as descriptive, understandable, and predictive from data [3]. One of data mining process is classification. It is a supervised learning technique to classifies the data items into a given class label [4]. There are many classification methods, such as; *Decision Tree, k-NN, Naive Bayes, SVM* and *Neural Network*. Each of methods is suitable for classify specific application. *Naive Bayes* and SVM normally used for *text classification*. *Decision Tree* is used to detect simple visualization pattern within complex variable data, *k-NN* could see the similarity of the data based on their respective distance, and *Neural Network* is able capture knowledge or predict event from complex data [5].

*k-NN* is one of the simple classification methods, the main idea of *k-NN* is measure the data similarity by using nearest neighbor distance of data [6]. *Decision Tree* is a model represented as two-way split binary tree to display the value of a target variable can be predicted by using the values of an independent variables [7]. Both are suitable for our e-commerce data characteristics.

Training data is an activity to find general pattern of the data to a target label. This activity is based on the data characteristics. Testing data is an activity to match data to the model resulted from training activity. This activity to determine model accuracy. As the result, the model is used to predict the any data value related to the model. We use *70:30* rules, where 70% of the data use for training activity and 30% of the data use for testing activity [8]. *Data training* and *data testing* should have a fair composition amount of data in each labeled target. Then, the data is evaluated by using confusion matrix, to see the object-based classification performance score [9].

## 3. Research Methods

We classify home sales data and create home sales model based on e-commerce historical sales data. The first step is to identify the problem, which is to define how to use the data to predict or to model home sales characteristics. The second step is set the research goal, which in this case the objective is the resulted model able to predict home sales based on their properties features. The third step is data collection process, where we collect only the useful data features to the model construction. The fourth step, after data has been collected, we do preprocessing activities to remove duplicate, incomplete, and other useless data. The fifth step is to divide data into 2 types; which are *Data Training* and *Data Testing*. The sixth step is model construction based on those 2 types of data. The last step is to evaluate the model accuracy. The complete workflow of our research methodology can be seen in Fig.1.

We collect the data by from e-commerce site *rumah123.com*. We get in total *801* home sales data in *Bandung*, from the 2015-2016 transaction period. The home sales data consists of price, building size, land size, bedroom, bathroom and location. We name the variable *building_size* to represent the building size, *land_size* represent land size, while *bedroom* and *bathroom* represent their respective variable. Example of home sales data can be seen in Fig.2

To simplify model classification construction, we define three numerical class as a transformation of continuous data value. This transformation is to simplify model, for example if we let the price range in continuous value, it will create large class to each different price value as the label target. This makes classification model become hard to understand. The three-numerical class in form of interval variables can be seen in Table 1. After we done preprocessing then, we split the data into training and testing data. 70% for data training and 30% data testing. After that, we construct the predictions model. Then, we evaluate the prediction model using confusion matrix tools.

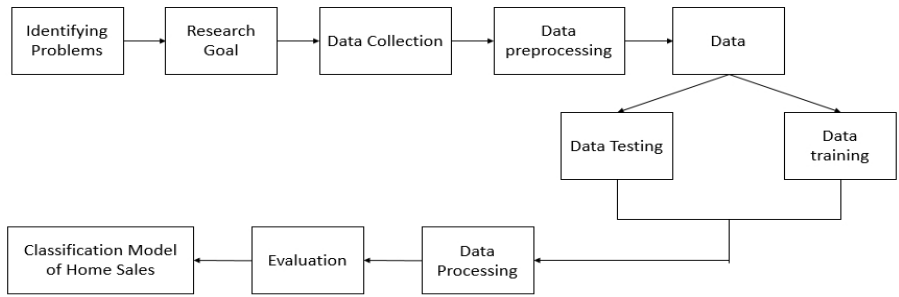

Figure 1 Research methodology

Figure 2 Example of home sales data in Bandung

Table I. Three numerical class of interval variables

| Price_Range (in Rp) | Land_Size (in m2) | Building_Size (in m2) | Price_Class |
|---|---|---|---|
| < 603500000 | < 107 | < 89 | Price_A |
| 603500000 - 1487500000 | 107 - 175.5 | 89 - 171 | Price_B |
| ≥ 1487500000 | ≥ 175.5 | ≥ 171 | Price_C |

## 4. Result and Analysis
## 4.1 Classification Model
4.1.1 Decision Tree

The decision tree home sales model is shown in Fig 3. The root node is *building_size* represent building size, while *land_size* represent land size as the leaf node. Both variable is significantly important variables to define property price. The variable bathrooms, bedroom and location are not significantly important in decision tree model. The reason for this is given by data characteristics of those three variables, which their value does not considerably varies across the data.

To understand the model, we give the following example; If we want to know what is the price for a property that have building size is 80m2 and land size is 100 m2. We trace the corresponding value in Node 1 and Node 5, then we conclude that the 91,5% probability that the price is fall into *Price_A* class, which is less than *Rp. 603500000*.

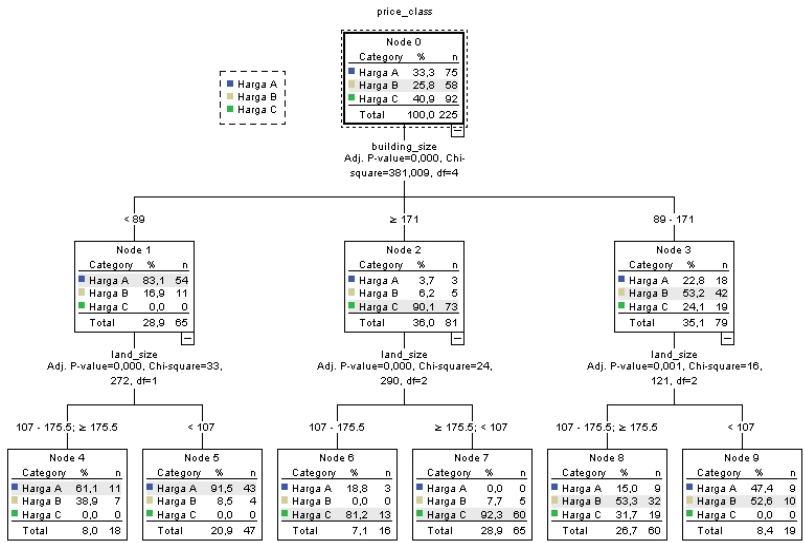

*Figure 2 Decision tree model*

### 4.1.2 k-Nearest Neighbor

The *k-NN* property classification model is shown in Fig 4. *k-NN* represent all the variables where it classifies the property data based on the data similarity or their closest distance across any available variables.

To understand the practicality of this model, we use following example; if we want to buy a home where the land size is 80 m2 land size and its building size is 80 m2. Our budget is not more than *Rp 1500000000*, -. We find data similarities on all the variables. There are 4 similar variables which are *land_size, building_size,* and *bathroom*. We have also 2 variables that mismatched which are *location* and *price_class*. We conclude that the decision fall into data in *Hegarmanah* with *Price_B* which is the price around between *Rp. 603500000* to *Rp 1487500000*.

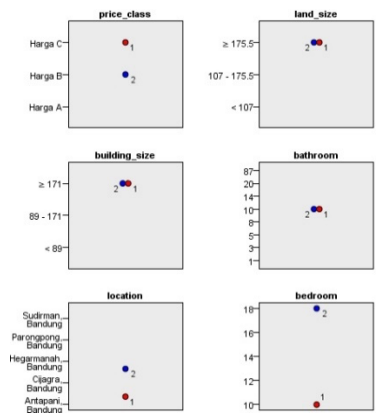

*Figure 3 k-NN classifications models*

### 4.1.3 Evaluation Accuracy Test

The evaluation or accuracy test is done by using *confusion matrix*. Table 2 show the evaluation of *Decision*

*Tree* and Table 3, show the evaluation of *k-NN* of home sales in Bandung. We have the *Decision Tree* accuracy is 75%, while *k-NN* is 71%.

*Table 2. Accuracy table of decision tree model*

| Partition | Observed | Predicted | | | |
|---|---|---|---|---|---|
| | | Harga A | Harga B | Harga C | Correct Percentage |
| Data | Harga A | 54 | 18 | 3 | 72,00% |
| | Harga B | 11 | 42 | 5 | 72,40% |
| | Harga C | 0 | 19 | 73 | 79,30% |
| | Overall | | | | 75,10% |

*Table 3. Accuracy table of k-NN model*

| Partition | Observed | Predicted | | | |
|---|---|---|---|---|---|
| | | Harga A | Harga B | Harga C | Correct Percentage |
| Data | Harga A | 53 | 11 | 3 | 79,10% |
| | Harga B | 14 | 48 | 10 | 66,70% |
| | Harga C | 2 | 17 | 45 | 70,30% |
| | Overall | | | | 71,90% |

## 4.2 Comparison Methodology

We sum up the difference of both methodology *Decision Tree* and *k-NN* model in this research, by the following comparison table in table 4.

*Table 4. Comparison of decision tree and k-NN model*

| *Model* | *Decision Tree* | *k-NN* |
|---|---|---|
| Data Source | Data from e-commerce *rumah123.com* | Data from e-commerce *rumah123.com* |
| Measurement Indicator | To predict the price, model is presented by *price_class* as target label. The most influential variables for the model are *building_size* and *land_size*. Other variables are not significantly important to predict the label target | To create exploration model, all variables are used to construct the model. Each data is measured their similarity by respective variables. |
| Analysis | *Decision Tree* produce prediction model | *k-NN* produce exploration model |
| Result of Accuracy | 75,10% | 71,90 % |

## 5. Conclusions

We compare two classification methods for home sales data in *Bandung*. Classification methods are used to construct the model for predict property prices. *Decision Tree* and *k-NN* has their own uniqueness. We found *Decision Tree* is more accurate than *K-NN* for given data characteristic. *Decision Tree* is suitable for prediction model while *k-NN* suitable for exploration model. For the future research, we suggest performing home sales data processing by using more specific attributes and others techniques like *unsupervised learning Clustering* to see a group in the data without label target.

### References

[1] Bhaskara, I. L. (2016, Desember 12). *Geliat e-Commerce Berkah Industri Logistik*. Diambil kembali dari tirto.id: https://tirto.id/geliat-e-commerce-berkah-industri-logistik-b854
[2] Hashem, I. A. T., Yaqoob, I., Anuar, N. B., Mokhtar, S., Gani, A., & Khan, S. U. (2015). *The rise of "big data" on cloud computing: Review and open research issues*. Information Systems, 47, 98-115.
[3] Zaki, M. J., & Meira Jr, W. (2014). *Data Mining and Analysis: Fundamental Concepts and Algorithms*. Cambridge University Press.

[4] Larose, Daniel T., and Larose, Chantal D. (2014). *Discovering Knowledge in Data: An Introduction to Data Mining Second Edition*. New Jersey: John Wiley & Sons Inc.
[5] Nikam S. S., (2015). *A Comparative Study of Classification Techniques in Data Mining Algorithms*. Oriental Of Computer Science and Technology.
[6] Parvin, H., Alizadeh H., Minati B. (2010). *A Modification on k-Nearest Neighbor Classifier*. Global Journal of Computer Science and Technology.
[7] Swamy, M, N., & Hanumanthappa, M. (2012). Predicting academic success from student enrolment data using decision tree technique. Int. K. Appl. Inf. Syst, 4, 1-6.
[8] Mucherino, A., Papajorgji J. P., Pardalos, P. M. (2008). *Data Mining in Agriculture*.
[9] Han, J. and Kamber, M, 2006, "*Data Mining Concepts and Techniques Second Edition*". Morgan Kauffman, San Francisco.
[10] H. P., Raden Johannes, and Alamsyah, Andry. (2015). *Sales Prediction Model Using Classification Decision Tree Approach for Small Medium Enterprise Based on Indonesia E-Commerce Data*. The 6[th] SCBTII.
[11] M.H., Handito, A., Citrananda, Alamsyah, Andry. (2016). *Predictions Models Based on Flight Tickets and Hotel Rooms Data Sales for Recommendation System in Online Travel Agent Business*. The 7[th] SCBTII.